# Artificial Eye Model and Holographic Display Based IOL simulator

Deniz Akyazi[a], Koray Kavakli[a], Ugur Aygun[a], Afsun Sahin[b], Hakan Urey*[a]
[a] Department of Electrical and Electronics Engineering, Koç University, Istanbul, Turkey, 34450
[b] School of Medicine, Koç University, Istanbul, Turkey, 34450

## ABSTRACT

Cataract is a common ophthalmic disease in which a cloudy area is formed in the lens of the eye and requires surgical removal and replacement of eye lens. Careful selection of the intraocular lens (IOL) is critical for the post-surgery satisfaction of the patient. Although there are various types of IOLs in the market with different properties, it is challenging for the patient to imagine how they will perceive the world after the surgery. We propose a novel holographic vision simulator which utilizes non-cataractous regions on eye lens to allow the cataract patients to experience post-operative visual acuity before surgery. Computer generated holography display technology enables to shape and steer the light beam through the relatively clear areas of the patient's lens. Another challenge for cataract surgeries is to match the right patient with the right IOL. To evaluate various IOLs, we developed an artificial human eye composed of a scleral lens, a glass retina, an iris, and a replaceable IOL holder. Next, we tested different IOLs (monofocal and multifocal) by capturing real-world scenes to demonstrate visual artifacts. Then, the artificial eye was implemented in the benchtop holographic simulator to evaluate various IOLs using different light sources and holographic contents.

## 1. INTRODUCTION

At least 2.2 billion people worldwide have vision impairment, of which 65.2 million have moderate to severe distance vision impairment or blindness due to cataracts [1]. These statistics suggest that cataract is one of the leading causes of vision impairment. Cataract treatment involves one-time surgery under local anesthesia, which can be performed as a day case. The surgery involves removing the opaque lens and implanting an intraocular lens (IOL) [2]. It is highly cost-effective [3] and significantly improves the quality of life [4] since surgery at an early stage can prevent the worsening of vision impairment or restore vision if undertaken later.

There are several options for the types of intraocular lenses to be implanted. A monofocal IOL with a single, fixed focal length has been a common choice for this implantation. Another option is multifocal IOLs, which are engineered to provide good unassisted distant and close vision. They realize multiple foci with diffractive optics, zones of varying refractive power, or their combinations [5]. If a multifocal IOL has two foci, it is called bifocal; if it has three foci, it is called trifocal.

Although they were engineered to have multiple foci for spectacle-free vision and have considerably good optical performance, in 7 trials, multifocal IOLs were related to undesirable visual phenomena in 97 of 334 instances (29%), compared to just 26 of 328 cases (8%) with monofocal IOLs [6]. The positive projection of non-useful patterns distinguishes these visual phenomena, also called pseudophakic photic phenomena, on the retina. These photic phenomena are artifacts that appear as glares or halos under certain lighting conditions [7]. Apart from these photic phenomena, patients with multifocal IOL implantations experience a decrease in contrast sensitivity, discontinuities in focus between the foci of the IOL and less resolution. As a result, the connection of multifocal IOLs with photic phenomena remains one of the most common reasons for postoperative complaints [6,8].

*hurey@ku.edu.tr, phone +90 536 569 40 96, https://mems.ku.edu.tr/

An objective evaluation and comparison of IOLs is a crucial help for patients and doctors in choosing which IOL to use. In this study, we combine holographic display technology with our newly designed optomechanical eye model and propose a novel benchtop setup for cataract patients. The developed platform predicted post-operative visual acuity with very good accuracy in 13 patients in clinical trials. Furthermore, to compare the performance of different IOLs we developed an artificial human eye with a replaceable IOL holder and tested it using real world objects to demonstrate side effects such as contrast decrease. Finally, we tested the artificial human eye using the holographic simulator to demonstrate the effect of the coherence properties of different light sources on IOLs.

In order to understand how the vision of patients change after surgery, real world scenes were captured using the artificial eye. Two IOLs were tested; (i) a monofocal IOL with 14.5 D optical power, (ii) a bifocal IOL with 15.5+3.25 D optical power. The image of a real-world scene taken with different intraocular lenses using the artificial eye model is given in Figure 4.

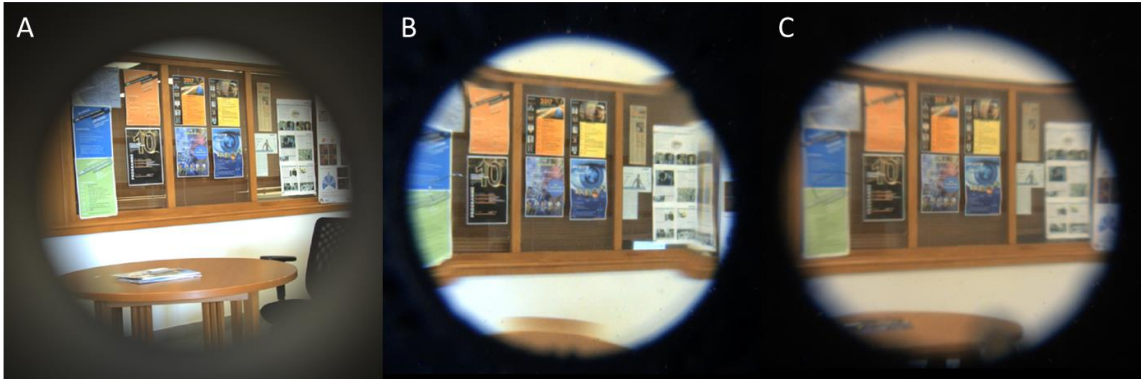

Figure 1. Posters on the wall were photographed using a camera with a standard lens. (A) Direct imaging of the scene with the camera. (B,C) Uses an artificial eye model using (B) monofocal IOL and (C) bifocal IOL to first form an image on a glass retina, which is then reimaged with the camera. While monofocal IOL nearly reached the image quality of (A), blur and contrast decrease can be observed in the multifocal IOL.

To further understand the problems occurring in a multifocal IOL vision, we observed a classroom scene with light sources. Captured images of the classroom where multiple light sources were present are given in Figure 2. Through multifocal IOL, the classroom is blurry compared to monofocal IOL and halos were observed around the light sources as well as a decrease in resolution even in focused planes.

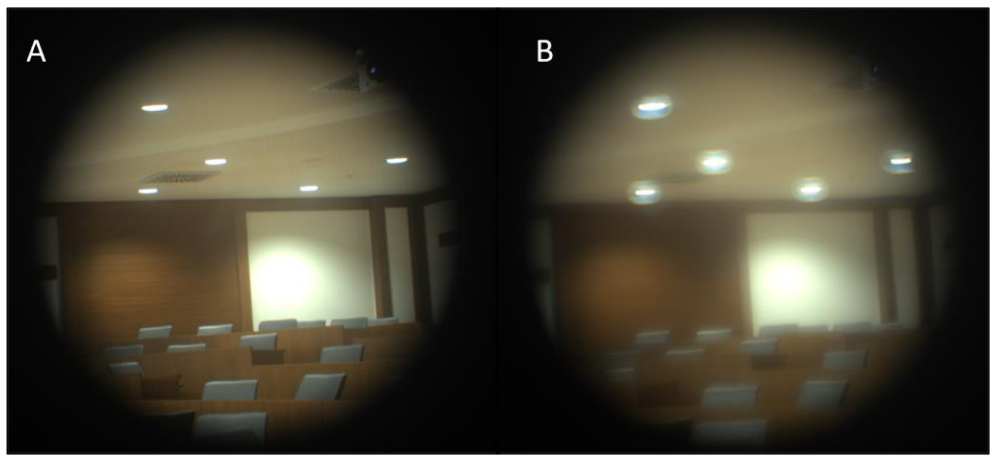

Figure 2. A classroom photographed with artificial eye model implanted with (A) monofocal IOL, (B) bifocal IOL, and camera. While monofocal IOL preserves the quality of vision, we start to observe contrast decrease, unfocused objects, and photic phenomena such as halos in bifocal IOL.

## 2. METHODOLOGY AND INSTRUMENTS

### 2.1 Optomechanical Eye Model and IOL Characteristics

We designed an artificial eye model based on a pseudophakic eye, to evaluate various IOLs and objectively test how the human eye would be affected with an IOL implanted. We first designed the optical system using Zemax optical design software to reach optimal optical properties. Since it was the most common multifocal IOL type, we took the bifocal IOL as the basis for modeling the intraocular lens in Zemax. Using two different configurations, we aimed for the resulting image to fall on the glass retina. For this purpose, the distance between the intraocular lens and the glass retina has been optimized, considering that the scleral and intraocular lenses should be as close to reality as possible.

In addition to optical criteria, we also considered mechanical design criteria. Since a system in which the eye is modeled is designed, we envisaged that the inner chamber of the design should be filled with a liquid with a refractive index close to that of the liquid inside the chamber. Considering the factors mentioned above, it was decided to produce the artificial eye model in a 3D printer, and the prototype was created. As seen from the spot diagrams, while bifocal IOL could focus two distances to the glass retina, monofocal IOL could only focus the far distance, this further validates our artificial eye model.

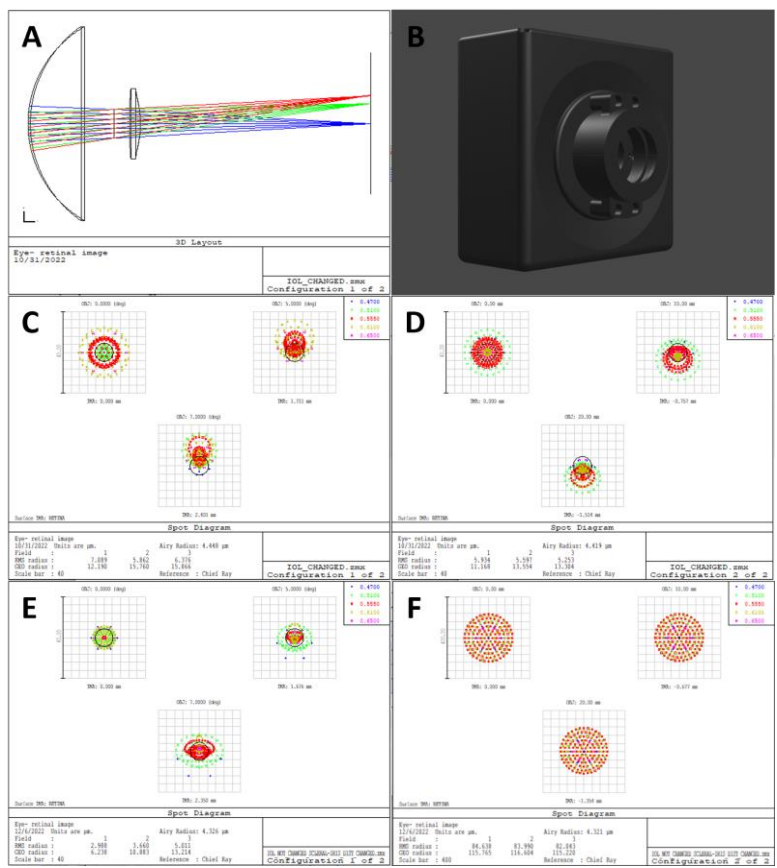

Figure 3. The artificial eye model in Zemax and 3D design program. (A) Layout for the artificial eye implanted with bifocal IOL. (B) Render of the design in Blender. Spot diagrams for two different IOL designs are given in (C-F). As can be seen while bifocal IOL can focus far distance (C) and near distance (D), monofocal IOL can only focus far distance (E) and an unfocus near distance (F).

Captured USAF Chart images for both lenses and their relative contrast plot with respect to distance are shown in Figure 5. As expected, bifocal IOL could focus at two different distances. On the other hand, contrast of the captured images through the bifocal lens was lower. When the contrasts plots are observed, we can see that the have a similar shape to the expected defocus curves of the IOLs. This validates our results further. While the best achieved visual acuity for monofocal IOL is around 20/22, it is 20/25 for the bifocal IOL.

Figure 4. Captures of IOLs in real life. First row is monofocal IOL captures for respectively near intermediate and far distance followed their contrast in lowest resolution. Second row is same captures and plots for bifocal IOL.

In addition to the resolution reduction, studies have also been conducted on the effect of contrast reduction due to intraocular lenses. For characterization, it was decided to extract the contrast functions of different lenses first. For this, USAF graph captures taken through artificial eye model with 35 cm distance were analyzed and the contrast percentage observed depending on the spatial frequency was calculated by averaging the areas. The results of the calculated contrast frequencies for monofocal and bifocal IOL were given in Figure 6. As expected, more contrast decrease in bifocal IOL was observed compared to the monofocal IOL.

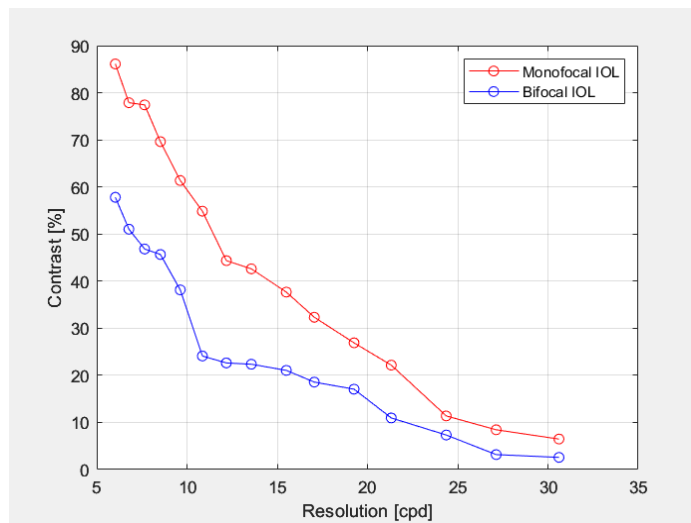

Figure 5. Contrast function of characterized IOLs through the artificial eye model. As seen on the plots, while monofocal IOL nearly preserves the contrast difference in low resolution, there is a decrease in bifocal IOL.

## 2.1 Holographic Display Setup

We developed a cataract simulator based on computer generated holography (CGH) display technology to allow cataract patients to experience post-operative visual acuity [1]. CGH display allows us to steer the light beam through the non-cataractous regions of the eye lens, hence patients can see images even through cataractous lenses. In this work we developed a benchtop version of our cataract simulator where we successfully predicted post-surgery visual acuity of patients. The holographic display consists of a light source, a spatial light modulator (SLM), 2 convex lenses, a Fourier filter, and a camera. It was decided to use a specially designed SLM as the SLM selection. The resolution of the SLM is 1920 × 1080 pixels and the pixel pitch is 8 µm. SLM provides phase modulation at 60 fps per second.

## 3. DATA AND RESULTS

The artificial eye model is integrated into the holographic vision simulator to test various IOLs. We used a phase-only liquid crystal on silicon (LCoS) spatial light modulator (SLM) to create holographic content. A He-Ne laser with a wavelength of 632.9 nm and an LED light source with a wavelength of 530 nm were used for illumination. Several different IOLs were tested using this implementation. Captured holographic images through different IOLs are given in Figure 3 for various focus planes and with different light sources. As can be seen from the figure, bifocal IOL has extended depth, however in the laser-based display case interference fringes are observed. These fringes are attributed to the coherence nature of the laser beams which deteriorate the captured images with multifocal IOLs.

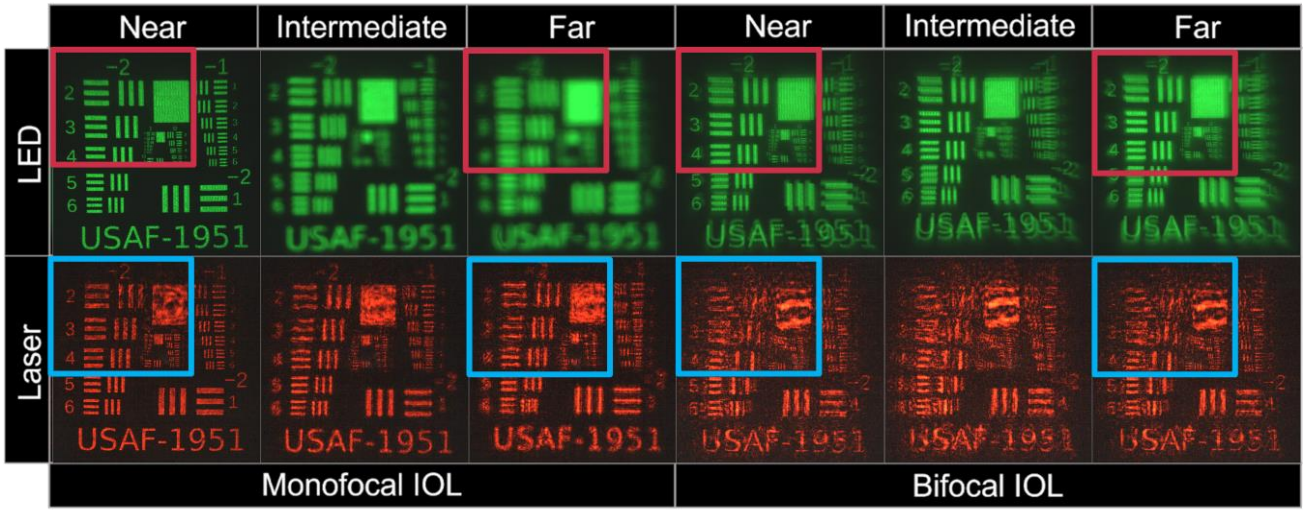

Figure 6. Captured holograms using artificial eye model with different IOLs and light sources. Upper row is LED illumination for near, intermediate, and far distances of Monofocal and Bifocal IOLs respectively, the lower row is laser illumination for same distances and IOLs.

## 4. CONCLUSION

Cataract is a common ophthalmic illness that causes a hazy region to form in the eye's lens and necessitates surgical removal and replacement of the eye lens. The intraocular lens must be carefully chosen to ensure the patient's pleasure after surgery. Even though there are many types of IOLs on the market with varying qualities, it is difficult for the patient to envision how they would view the world following the surgery. We propose an artificial human eye with a scleral lens, a glass retina, an iris, and a removable IOL holder to test different IOLs and characterize their effects to vision and their visual artifacts. To exhibit these effects and visual artifacts, we evaluated several IOLs (monofocal and multifocal) by collecting real-world situations. We then developed a holographic vision simulator that uses non-cataractous portions of the eye lens to provide cataract patients with pre-operative visual acuity. Shape and direct the light beam via the comparatively clear portions of the patient's lens using computer produced holography display technology. The artificial eye was then implemented in the benchtop holographic simulator to assess multiple IOLs with varying light sources and holographic contents. In the end, we successfully characterized various chosen IOLs under different light conditions and real scenes. Further research will be conducted to detail the characterizations and apply them to holographic contents.

## ACKNOWLEDGEMENT

The authors would like to thank Erdem Ulusoy and Bora Carpinlioglu for the help in holographic content development. The authors also thank Firat Turkkal for helping with the artificial eye model in computer-aided design and drafting. This work has been supported by TUBITAK (grant agreement number 120C145) and European Innovation Council (EIC) under grant agreement no 101057672.